# Applying Autonomy with Bandwidth Allocation Models

Rafael Freitas Reale, Romildo Martins da S. Bezerra
DMCC / INSERT
Federal University of Bahia / Federal Institute of Bahia
Salvador, Bahia, Brazil
reale@ifba.edu.br, romildo@ifba.edu.br

Joberto S. B. Martins
NUPERC
Salvador University
Salvador, Bahia, Brazil
joberto.matins@unifacs.br

*Abstract* — Bandwidth Allocation Models (BAMs) are resource allocation methods used for networks in general. BAMs are currently applied for handling resources such as bandwidth allocation in MPLS DS-TE networks (LSP setup). In general, BAMs defines resource restrictions by "class" and allocate the available resources on demand. This is frequently necessary to manage large and complex systems like routing networks. G-BAM is a new generalized BAM that, by configuration, incorporates the "behavior" of existing BAMs (MAM, RDM, G-RDM and AllocTC-Sharing). In effect, any current available BAM "behavior" is reproduced by G-BAM by simply adjusting its configuration parameters. This paper focuses on investigating the applicability of using autonomy together with Bandwidth Allocation Models (BAMs) for improve performance and facilitating the management of MPLS DS-TE networks. It is investigated the applicability of "BAM switching" using a framework with autonomic characteristics. In brief, it is investigated the switching among "BAM behaviors" and BAM´s reconfiguration with distinct network traffic scenarios by using G-BAM. Simulation results suggest that the autonomic switching of "BAM behaviors" based on high-level management rules (SLAs, QoS or other police) may result in improving overall network management and operational parameters such as link utilization and preemption.

## I. Motivation

Bandwidth Allocation Models (BAMs) are resource allocation methods used for networks in general. The handling of on-demand resource allocation such as bandwidth (LSP - Label Switched Path setup) in MPLS DS-TE networks is the main target of BAMs. From the management point of view, BAMs need high-level administration including manager's specialization and constant monitoring of the network traffic and operational parameters in order to adopt the most suitable among the currently available BAM models: MAM [1], RDM [2], G-RDM [3] or AllocTC-Sharing [4] .

The choice for a model (BAM) and its parameters configuration directly imply in complying with the dynamicity of input traffic, defined SLAs (Service Level Agreement) and other managerial high-level requirements. The decision about the most suitable BAM also depends on the actual state of the network. In brief, managing BAM adequacy for networks implies in deciding on the most suitable BAM to adopt followed by configuring it according with the actual network traffic demand. Typically, a long search time is required for computing the BAM solution for problems and, beyond that, it needs a high-level of specialization for managers and constant network monitoring. [5]. An autonomic choice for BAM model and related parameters configuration is then the main motivation of the investigation presented in this paper. To the extent of our knowledge, the literature has not yet focused on this kind of approach whether autonomic decisions are proposed to both choose and configure an specific BAM according with the actual network state. The approach adopted is in the direction to allow dynamic BAM switching based on new G-BAM model and consider the dynamic network traffic demand [6].

In general, Network Management Systems (NMS) need more automated and efficient management solutions for current networks like IP/MPLS, optical networks and Next Generation Network (NGN) [7] [8]. Figure 1 illustrates the basic autonomic management levels and its evolution. There is an evolutionary process towards a fully autonomic solution motivated mainly by the development of the technology and increasing complexity of actual systems. The fully autonomic management solution represents the final management levels in which some reasoning and learning approaches are included (Fig. 1).

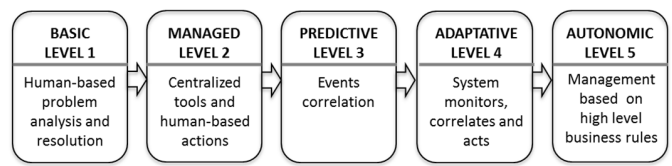

Fig. 1. Management evolution levels towards autonomy [9] [10]

This paper aims to investigate the application of autonomic management together with BAMs. The target is to evaluate the applicability and define basic requirements for the steps involved in the autonomic computation, including monitoring, analysis, planning and execution. The expected overall result is to determine the viability to manage with autonomic characteristics the dynamic behavior of BAM-based networks supporting an on-the-fly operation towards network optimization.

In brief, it is investigated the switching among "BAM behaviors" and BAM´s reconfiguration with distinct network traffic scenarios by using G-BAM.

The structure of this paper is as follows: Section II presents a summary with the main characteristics of the most adopted Bandwidth Allocation Models (MAM, RDM, G-RDM and AllocTC-Sharing) according to their allocation characteristics and resources sharing. G-BAM is then presented with emphasis on its capability to generalize all other BAM´s "behaviors". Section III explores the application of autonomy with BAMs. Autonomic management implementation steps (monitoring, analysis, planning and execution) by using a framework are considered together with the inherent "behavior" of BAM models. Finally, section IV presents a case study scenario oriented to the evaluation of BAM with autonomic characteristics.

## II. BANDWIDTH ALLOCATION MODELS – AN OVERVIEW

Bandwidth Allocation Models (BAMs) were originally proposed to define rules and limits (Bandwidth Constraints - BCs) for links usage by traffic classes (TCs) in DS-TE (MPLS Traffic Engineering) networks. The BAMs effectively define how resource (bandwidth) is obtained and shared among traffic classes (applications). Operationally, the application of a BAM in MPLS DS/TE networks results in setup, blocking or preemption of LSPs for specific links.

MAM, the most basic model, aims to support network traffic profiles in which a strong isolation between traffic classes (TCs) is required. In MAM, traffic classes use only private resources and there is no resource sharing (bandwidth) between traffic classes.

RDM allows non-allocated resources (bandwidth) belonging to high priority TCs to be shared by low priorities TCs. Preemption of less priority LSPs using shared resources is required by high priority LSP setup (priority applications) in order to maintain defined bandwidth constraints (BCs).

AllocTC-Sharing model extends the strategy of non-allocated resources sharing to high priority TCs. As such, high priority TCs (applications) may "loan" (share) unused resources belonging to low priority TCs. A loan corresponds to a temporary use of the bandwidth allocated in low priority TCs by LSPs belonging to a high priority TC. In AllocTC-Sharing, "devolution" is the teardown of an established priority LSP using shared resources by loan.

G-RDM is a hybrid model between RDM and MAM. G-RDM uses private and sharable resources. As such, it incorporates both MAM and RDM behaviors.

### A. G-BAM – A Generalized Bandwidth Allocation Model

The Generalized Bandwidth Allocation Model (G-BAM) is an innovative bandwidth allocation model that allows the definition of private resources, incorporates all sharing resources alternatives (HTL[1] and LTH[2]) and, beyond all this, has the capability to reproduce the "behavior" of other proposed BAMs, such as MAM, RDM, G-RDM and AllocTC-Sharing [6]. The new G-BAM model reproduces these "BAM behaviors" by reconfiguring its operational parameters.

Among the new possible allocation strategies provided by G-BAM, it is worth to mention:

- The integration of private resources with AllocTC-Sharing behavior (HTL + LTH); and
- The integration of private resources with LTH.

An important advantage in adopting G-BAM, instead of a single BAM or multi-BAMs for a network is the possibility to provide to the network administrators a reconfigurable BAM solution. G-BAM allows the possibility to adapt the currently adopted BAM behavior to distinct traffic profiles.

Actually, G-BAM has the full capability to deal with the dynamics of the network traffic profile by making use of its configurable "BAM behavior". Its configurable BAM behavior, beyond the possibility to switch between distinct BAMs, also allows the configuration of intermediary and transitory network states (behaviors). Transitory and intermediary behaviors are not possible to setup when using the strategy of switching between existing BAM models (MAM to RDM or RDM to AllocTC-Sharing, other switching alternatives) [11].

Another relevant aspect of BAM switching is the migration strategy adopted [12]. Since G-BAM is a single configurable model, sharing values for HTL, LTH and private resources can be adjusted in order to provide both "SOFT" or "HARD" migration between BAM behaviors. Operationally, G-BAM also supports the reconfiguration of BCs.

The above-mentioned set of G-BAM capabilities provides additional implementation methods that can be used to support new classes and dynamic traffic profile configurations that are not supported with basic BAMs (MAM, RDM and AllocTC-Sharing) in a single, multi-BAM, static and/or dynamic BAM switching implementation.

As such, the following section explores the application of autonomy with BAMs considering G-BAM as the model adopted by the network.

## III. APPLYNG AUTONOMY WITH BAMS – BASIC APPROACHES ADOPTED

Applying autonomy with BAMs means, fundamentally, to look for the most suitable BAM and related configuration in order to achieve high-level management rules, like SLAs and other high-level business and management constraints, under the condition of dynamic input traffic.

In brief, this corresponds to a specific case of resource allocation with BAMs being defined as the basic mechanism adopted to control resources. This is similar to the resource allocation problem found in cloud computing [13], traffic engineering [14], datacenters [15] and access and wireless networks [16], where technologies, algorithms, methods and/or heuristics are used.

---

[1] HTL: high-to-low sharing – non-allocated resources belonging to high priority TCs can be shared by low priority TCs.

[2] LTH: low-to-high sharing: non-allocated resources belonging to low priority TCs can be shared by high priority TCs.

Applying autonomy with BAMs will consider three strategies in this paper:

- Dynamic BAM switching;
- BAM´s parameter dynamic reconfiguration; and
- BAM optimization.

### A. Dynamic BAM switching

In effect, each possible BAM (MAM, RDM, AllocTC-Sharing or other) adopted/ configured in a network has a specific behavioral characteristics (BAM "behavior") that will result indifferent sets of operational parameters (utilization, blocking, preemption, …) for the same input traffic profile. In other words, link utilization, preemptions, blocked LSPs and devolutions will change for the same input traffic for distinct BAMs adopted.

Therefore, BAM dynamic switching (an on-the-fly BAM switching approach) is effectively a method to achieve a defined network state and/or control operational parameters as required by the network administration. As a simple example, the number of preemptions may be kept below a defined limit by switching BAM "behaviors" (MAM to RDM or to AllocTC-Sharing) as a function of the network input traffic.

Autonomy may play a key role in this operational scenario since there is a potential decision whether to change or not among BAM models. Another way to perceive this strategy is to take into account that link utilization, preemptions, blocked LSPs (blocking) and devolutions are some operational parameters that can be autonomously managed/controlled by choosing the BAM to be configured as the resource (bandwidth) controller.

### B. Reconfiguration of BAM´s Bandwidth Constraints (BCs)

Obviously, any BAM "behavior" strongly depends on its configuration parameters like, for instance, the BCs (Bandwidth Constraints) defined for each traffic class (TC). As such, another set of possibilities to implement autonomic management with BAM´s support relies on changing their BC configuration.

However, the basic approach frequently used by managers for BAMs is to keep BAM´s BC parameters unchanged (static) independently of input traffic for all traffic classes. This is an "understandable" natural approach adopted by most network managers because the computation of a new network state (represented by a set of operational parameters like utilization, blocking and others) does require a lot of computation time and requires some expertise from previous network states. As such, the "standard" approach is to calculate the best BAM "behavior" for a specific medium or commonly expected traffic profile and adopt this configuration statically.

The computer networks are dynamic, asymmetric, multi-path and with limited resources in terms of bandwidth. Given this, an autonomic management approach may explore the reconfiguration of bandwidth constraints (BCs) for any adopted BAM in order to comply with high-level requirements defined for network operation.

In [11], two preliminary approaches for BAM´s BC reconfiguration were proposed and evaluated for distinct traffic profiles: (i) the "SOFT" approach; and (ii) the "HARD" approach.

The "HARD" approach premise is to reconfigure BCs immediately whenever BAM reconfiguration is decided to be necessary according to network traffic profile.

The "SOFT" approach premise is to reconfigure BCs slowly and adopts a "transition time" between actual and future network state.

For the three allocation models analyzed in [11] (MAM, RDM and AlloctTC-Sharing) the "HARD" approach leads to a new state at the cost of having "forced" preemptions resulting from the fact that the availability of resources has changed. As such, the drawback is that some applications pay the price to get the network to a new and, expectably, more "efficient" state. Anyway, this is an issue to consider with any autonomic solution.

The "SOFT" solution minimizes the operational impact at the cost of requiring a management plan to coordinate the reconfiguration transition time. In this case, the migration between network states is executed gradually. The operational effect of this approach is to better preserve the BAM´s operational characteristics at the cost of a variable convergence time to the new network state.

### C. BAM optimization

Another possibility to apply autonomy with BAMs is to consider optimization in general for a specific BAM. In this case, it is a matter of evaluating Bandwidth Constraints (BCs) and priorities among Traffic Classes (TCs) and TC application and services mapping for a single BAM in order to find out the best possible configuration in terms of high-level constraints (SLA, QoS, other). This is a multi-objective optimization problem involving a high number of variables and constraints.

## IV. APPLYING AUTONOMY WITH BAMs – THE FRAMEWORK

The implementation of autonomy with BAMs makes use of a framework that follows the basic control loop adopted by most autonomic management systems [10] [17]. The BAM framework with autonomic characteristics is based on three planes corresponding to the management cycle (Fig. 2):

- Information Plane: that monitors the network in relation to the set of monitoring parameters adopted by the BAM autonomic solution.
- Knowledge Plane: that analyses network monitored data and plans for effective actions.
- Execution Plane: that acts on the BAMs and network configuring or switching them according with the actions defined by the knowledge plan.

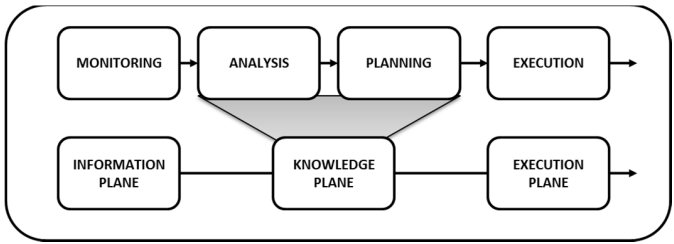

Fig. 2. BAM autonomic framework and the management cycle

The challenge involved in implementing the BAM autonomic framework is mostly concentrated on acquiring (on-the-fly computing) the knowledge supporting the decisions to either switch among BAMs or reconfigure them in order to comply with high-level management requirements.

The BAM framework adopts a centralized operation for monitoring data collection, knowledge computation and execution dispatch.

### A. Monitoring

The monitoring phase of the BAM autonomic framework aims to collect (gather) data elements that will be filtered and aggregated in order to be used at analysis phase.

The basic parameters (metrics) defined for the monitoring phase in the context of BAM´s autonomic management are:

- Utilization (Utilization) – corresponds to the effective bandwidth used in relation to link capacity for all network links.

- Number of Preemptions (Preemption) – the total number of accumulated "preemptions" in the network (MPLS DS-TE) in a configured period. The network manager configures the evaluation period.

- Number of blocked LSPs setup (Blocking) – A "blocking" is computed every time an LSP request cannot be granted in the network.

- Number of "Devolutions" (Devolution) – corresponds to the premature teardown of a priority LSPs established by "loan".

- Number of "Loans" (Loan) – The "loan" corresponds to the temporary use of bandwidth allocated to less priority TCs (LSPs) with this resource (bandwidth) belonging to higher priority TCs (LSPs).

- Number of LSPs Request (Request) – Number of requests for new LSP´s setup.

These are performance metrics resultant from the utilization of different BAMs collected by link and by TC. Effectively, BAMs act at link level arbitrating the concurrency for resources (bandwidth) between TCs. Metrics are collected and aggregated in periods defined by the network manager (day/night, week/weekend, holidays and others).

### B. Knowledge Plane (Analysis and Planning)

The "analysis phase" of the autonomic knowledge management cycle determines the current network state and possible future transitions based on the monitoring metrics gathered. Generally, this represents a complex analysis using various techniques in order to identify the several possible states and determine the next state that complies with the high level management requirements (Fig 3).

The analysis phase of the BAM autonomic framework, as a first implementation approach, determines three possible actions to be planned by the next phase (planning phase):

- BAM Switching action – results from an analysis concluding that BAM switching is the best possible recommended action for the network;

- BAM Reconfiguration action – results from an analysis concluding that BAM reconfiguration is the best possible recommended action for the network.; and

- BAM Optimization action – results from an analysis concluding that actual BAM may be kept, but overall network behavior can be improved by adjusting operational parameters of the BAM in use.

As previously indicated, each bandwidth allocation model has specific characteristics that result on distinct network operational parameters (link utilization, preemption, blocking, devolution and others) as a function of the input traffic. Thus, the "analysis phase" typically may decide on the most suitable bandwidth allocation model or recommend a new BAM configuration.

Obviously, BAM´s behavior (MAM, RDM and AllocTC-Sharing) have already been identified or are deductible from its overall characteristics. The autonomic framework analysis phase is expected to "perceive" the scenarios that can be further verified by simulation.

Some more obvious "perceptions" are illustrated next for MAM, RDM and AllocTC-Sharing.

MAM model is indicated when the administrator expects a behavior in which a strong isolation between Traffic Classes (TC) is required. The MAM model typically implies in low utilization of links and a high number of blocking since the available resources of a TC cannot, by definition, be shared with other TCs (total isolation of TCs). On the other hand, by not allowing the sharing of TCs, MAM model does not generate preemptions and devolutions.

RDM model is indicated when the administrator expects that the resources not used by high priority TCs can shared the low priority TCs. With more bandwidth available for low priority TCs the positive impact is higher link utilization and lower probability of blocking. The high priority bandwidth used to benefit low priority TCs can be requested back generating preemptions.

AllocTC-Sharing model supports the sharing of available bandwidth of high priority TCs for low priority TCs and vice-versa. This "opportunistic" model maximizes link utilization and minimizes the probability of blocking when allowing the utilization of the available link bandwidth. The monitoring of the levels of preemption and devolution is important for this model since LSPs and TCs owners of the shared bandwidth can request them resulting in preemption and/or devolution.

In general, some behaviors become evident in terms of the analysis when links tend to saturate, such as:

- The more sharing, the better link utilization and possibly smaller blocking;
- The more utilization of high to low sharing, the higher the probability of preemptions;
- The higher utilization of low to high sharing, the higher the probability of devolutions;
- In a network frequently experiencing congestion, models with conservative politicies are, in principle, the best solution (e.g. MAM or its equivalent configuration); and
- In a network frequently experiencing unsaturated links, models with more flexible politicies can be the adopted (e.g. AllocTC-Sharing or its equivalent configuration).

The "planning phase" of the knowledge plane receives state information and network state transition (decision/action) from the "analysis phase" aiming to accomplish the specified management objective. The computed action can affect only one or several managed elements in the network.

The "planning" defined within the BAM autonomic framework has, as a first implementation approach, three possible actions (Fig 3): BAM switching action; BAM reconfiguration; and BAM optimization action.

"BAM Switching" in terms of the planning phase consists in planning (computing) a soft or hard transition. "BAM Reconfiguration" in terms of the planning phase consists in planning (computing) a new configuration of BCs. The "BAM Optimization" phase corresponds to the computation of a new configuration setup for the current BAM and subsequent application to the network by the execution plane.

The analysis and planning cycle is repeated according the period configured by the administrator or in case a new network state is received by the knowledge plan.

*C. Execution Plane*

The execution phase controls the set of actions defined and planned in the previous phases. In short, execution phase transforms into practical actions the result obtained from previous phases. This change affects routers that can have different form of access and commands.

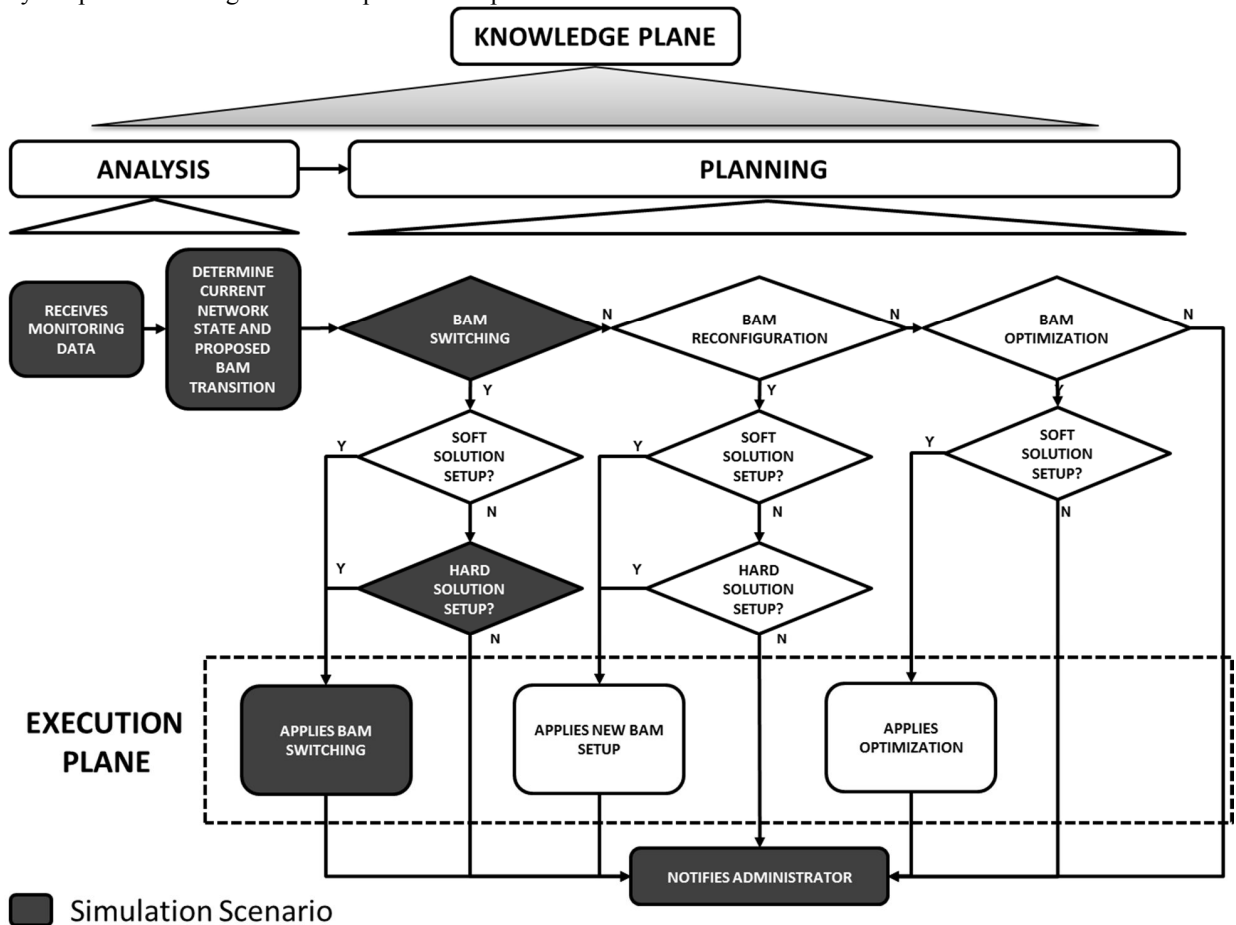

Fig. 3. The analysis and planning phase – BAM autonomic framework

## V. SIMULATION SCENARIO

The main purpose of this evaluation is to demonstrate the applicability and need of autonomy with bandwidth allocation models in IP networks.

The BAM used in the evaluation is the G-BAM. In the simulation performed, G-BAM is switched between two "behaviors": MAM and RDM behaviors. The BAM switching simulated adopts the HARD approach and explores the concept of autonomy applied to MAM e RDM "behaviors" as presented in previous sections.

G-BAM configuration for MAM and RDM behaviors are the reference for comparison with a new solution with autonomic characteristics. The simulation implemented in this scenario alters dynamically the configuration of G-BAM in order to switch from MAM to RDM and vice-versa. It is based on three intuitive parameters configurable by the network administrator: Observation window; Maximum number of preemptions; and Link utilization.

The specialized bandwidth allocation model simulator BAMSim [4] is used for the simulation. BAMSim has a central managing unit that has a full view of network´s topology (Fig. 4). BAMSim receives a routing plan and configuration (routing paths) that is kept during all the simulation scenarios.

The network topology used is the NTT network containing 55 nodes and 144 links of 622Mbps (STM-4 – SDH). For this evaluation scenario, we defined a single traffic source, in the router 0, for each traffic class and destiny (Fig. 4). The principle is that the simulation is evaluating the applicability of BAM with autonomy and, as such, behaviors can be extracted from any set of input traffic.

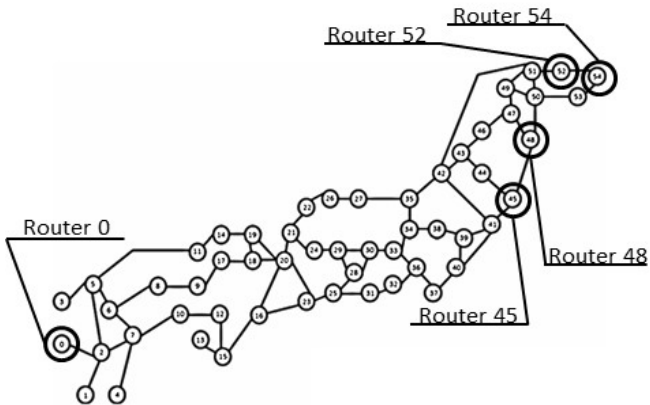

Fig. 4. Network topology used in the simulation

The (randomly) chosen destinations adopted in the simulation were routers 54, 52, 48 and 45 and three traffic classes are configured: TC0 - low priority applications; TC1 - medium priority applications; and TC2 - high priority applications.

G-BAM was configured reflecting bandwidth constraints (BCs) according to Table I.

TABLE I. BANDWIDTH CONSTRAINT BY TRAFFIC CLASS (TC)

| BCs | Max BC (%) | MAX BC (Mbps) | TC per BC | Max BC (%) | MAX BC (Mbps) | TC per BC |
|---|---|---|---|---|---|---|
| | | G-BAM (RDM) | | | G-BAM (MAM) | |
| BC0 | 100 | 622 | TC0+TC1+TC2 | 40 | 248,8 | TC0 |
| BC1 | 60 | 373,2 | TC1+TC2 | 30 | 186,6 | TC1 |
| BC2 | 30 | 186,6 | TC2 | 30 | 186,6 | TC2 |

Routes (LSP paths) are statically defined in order to have competition in a high number of links. Consequently, saturated links are forced during simulation in order to observe the consequences:

- 0->2->5->11->14->19->20->21->22->26->27->35->42->51->**52**
- 0->2->5->11->14->19->20->21->22->26->27->35->42->51->52->**54**
- 0->2->5->11->14->19->20->21->22->26->27->35->42->41->**45**
- 0->2->5->11->14->19->20->21->22->26->27->35->42->41->45->**48**

The configuration parameters for LSPs are:

- LSPs length exponentially modeled establishment – average of 200 seconds, leads to link saturation;
- LSP bandwidth – uniformly distributed between 5 Mbps and 15 Mbps;
- Halting criteria – 1 hour (3600s); and
- Interval of LSPs requests arrival modeled exponentially in phases according the table below:

TABLE II – RATE OF LSP ARRIVALS BY TRAFFIC CLASSES (TCs)

| Phase | 1 | 2 | 3 | 4 | 5 | 6 | 7 | 8 |
|---|---|---|---|---|---|---|---|---|
| Time (s) | 300 | 600 | 900 | 1500 | 1800 | 2100 | 2500 | 3600 |
| TC0 | 8 | 8 | 8 | 8 | 8 | 8 | 8 | 8 |
| TC1 | 0 | 8 | 8 | 8 | 100 | 100 | 8 | 50 |
| TC2 | 0 | 0 | 8 | 100 | 100 | 8 | 8 | 50 |

Phases 1 to 3 create a traffic profile where there are, initially, only low priority LSPs, then medium priority LSPs come and, finally, they are followed by high priority LSPs in a high flow rate forcing them to be used in the maximum. On phase 4, we reduced the rate of high priority LSP arrivals and, on phase 5, the medium priority ones. On phase 6, we maintained a low arrival rate of medium priority LSPs and we increased the arrival rate of high priority LSP. On phase 7, we generate a high number of LSPs for all classes in order to saturate the link. Finally, on phase 8, we reduced the arrivals of high and medium priority LSPs and we maintain the high arrival flux of low priority LSPs.

## VI. EVALUATION

The number of preemptions and link utilization are the first set of parameters considered in the evaluation towards the applicability of autonomy with BAMs.

Figure 05 illustrates the MAM behavior (configured with G-BAM) in which there are no preemptions (inherent behavior of MAM) but implies in links not being fully used for the entire simulation window. In other words, link utilization most of the time is below the 622Mbps TC capability even when TC0, TC1 and TC2 are saturated.

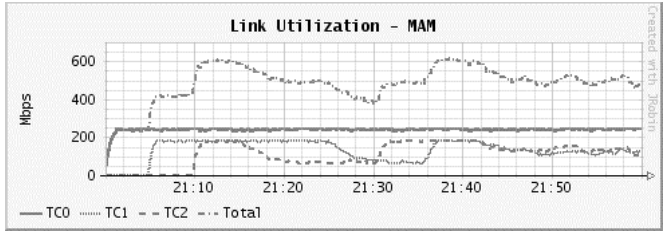

Fig. 5. Link utilization with G-BAM´s MAM behavior

Figures 06 and 07 illustrate the RDM behavior (configured with G-BAM). In this case, it is observed that link utilization is fully maximized being at the maximum capability of 622 Mbps most of the time. The drawback in keeping a high link utilization with RDM behavior is that it also results in a high and cumulative number of preemptions (Figure 7).

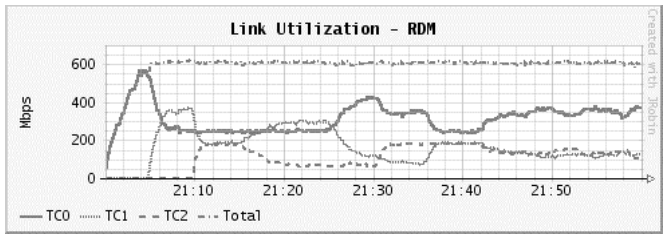

Fig. 6. Link 0 "utilization" (router 0 to 2) with G-BAM´s RDM behavior

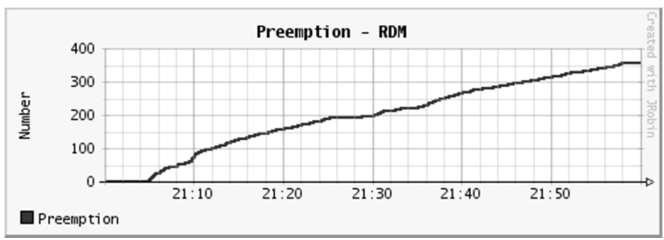

Fig. 7. Preemptions using G-BAM´s RDM behavior

As such, the simulation, until this point, indicates that there is a potential decision between the utilization of these two BAM models/"behaviors". As a very simple example, it could be required by the high-level management that link utilization should be maximized but the acceptable number of preemptions should be limited to a certain value/ operational parameter.

In terms of the application of autonomy with BAMs, this effectively means that BAMs should be either switched or reconfigured or optimized, as illustrated in Figure 03 (knowledge planning phase of the management process with autonomic characteristics).

In order to illustrate more clearly the BAM switching alternative (one out of three possible options to autonomic management) and the applicability of autonomy for the resulting possibilities, a specific simulation was realized in which:

- RDM is switched to MAM using the HARD approach when the number of cumulative preemptions reaches a certain level (this prevents further preemption to happen).
- MAM is switched to RDM using the HARD approach when link utilization reaches a certain level (this attempts to improve link utilization).
- Various (arbitrarily defined) high-level management parameters (number of preemptions/ link utilization tuple on Table III top line) are statically defined and the simulated results (means) are summarized in Table III:
    o Maximum number of preemptions in the defined time window;
    o Link utilization in the defined time window.
- Figures 08 and 09 illustrate the simulation results for a set of high-level configuration parameters as follows:
    o RDM to MAM switching with 25 cumulative preemptions and MAM to RDM switching with link utilization below 65% (considering the defined window).

Eight simulations were run with 05 random seeds. The results are presented in terms of the mean value obtained in Figures 08, 09 and Table III. The 5 minutes observation window in the simulation is a manager defined configuration parameter. RDM to MAM switching using HARD approach generates additional LSP preemptions in order to liberate shared bandwidth (MAM inherent operation and configuration).

TABLE III – SIMULATION RESULTS (MEANS) WITH PREEMPTIONS/ LINK UTILIZATION TUPLE ARBITRARILY DEFINING BAM SWITCHING

| | Simulation of Autonomic Characteristics | | | | | | | | |
|---|---|---|---|---|---|---|---|---|---|
| MAM | 25/65 | 30/65 | 25/70 | 30/70 | 25/75 | 30/75 | 25/80 | 30/80 | RDM |
| Number of Preemptions | | | | | | | | | |
| 0 | 105 | 122 | 125 | 140 | 154 | 179 | 184 | 209 | 358 |
| Blocking | | | | | | | | | |
| 1882 | 1687 | 1679 | 1655 | 1635 | 1608 | 1591 | 1580 | 1543 | 1335 |
| Generated LSPs | | | | | | | | | |
| 3269 | 3269 | 3269 | 3269 | 3269 | 3269 | 3269 | 3269 | 3269 | 3269 |
| Established LSPs | | | | | | | | | |
| 1387 | 1582 | 1590 | 1614 | 1634 | 1661 | 1678 | 1689 | 1726 | 1934 |
| LSPs not Preempted | | | | | | | | | |
| 1321 | 1406 | 1398 | 1415 | 1419 | 1428 | 1418 | 1481 | 1437 | 1494 |
| Generated LSP (Bandwidth) | | | | | | | | | |
| 26208 | 26208 | 26208 | 26208 | 26208 | 26208 | 26208 | 26208 | 26208 | 26208 |
| Established LSP (Bandwidth) | | | | | | | | | |
| 8639 | 9517 | 9494 | 9681 | 9701 | 9804 | 9763 | 9843 | 9940 | 10549 |
| Blocking (CT0) | | | | | | | | | |
| 1081 | 905 | 892 | 877 | 857 | 841 | 822 | 819 | 786 | 664 |
| Blocking (CT1) | | | | | | | | | |
| 485 | 467 | 470 | 461 | 462 | 450 | 453 | 445 | 441 | 355 |
| Blocking (CT2) | | | | | | | | | |
| 316 | 316 | 316 | 316 | 316 | 316 | 316 | 316 | 316 | 316 |
| Number of Preemptions (TC0) | | | | | | | | | |
| 0 | 93 | 104 | 110 | 117 | 130 | 149 | 152 | 170 | 288 |
| Number of Preemptions (TC1) | | | | | | | | | |
| 0 | 11 | 14 | 15 | 23 | 24 | 30 | 33 | 39 | 70 |
| Number of Preemptions (TC2) | | | | | | | | | |
| 0 | 0 | 0 | 0 | 0 | 0 | 0 | 0 | 0 | 0 |

Table III shows that the preemption values in the 05 minutes observation window and the maximum utilization of the bottleneck link have various intermediary behaviors between MAM and RDM.

Again, in terms of the applicability of autonomy in the BAM switching, configuration or optimization context, it is show in Table III that various "management decisions" are possible and, as such, that could be supported by a framework with autonomic characteristics.

As a simple example of autonomic management decision, the tuple 25/65 (maximum number of 25 preemptions/ link utilization of 65%) simulation is illustrated in Figures 08 and 09. In this case, G-BAM switches from RDM to MAM behavior when preemption limit is reached and, as such, keeps preemption stable (fixed). The cost is that link utilization is reduced. In the same way, an autonomic decision to switch from MAM to RDM is executed when link utilization falls below 65%. In this case, preemption might increment again. The point is that these simulations point out to the applicability of BAMs with autonomic characteristics in order to decide and/or follow transitions that are the most adequate according with the defined high level management configuration requirements.

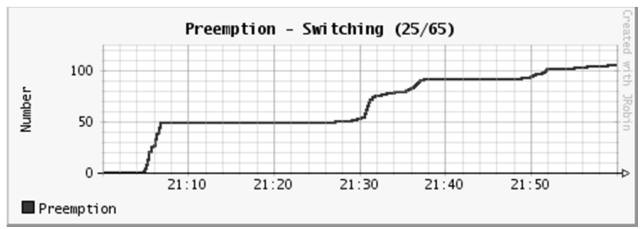

Fig. 8. Preemptions with dynamic switching configured as 25/65

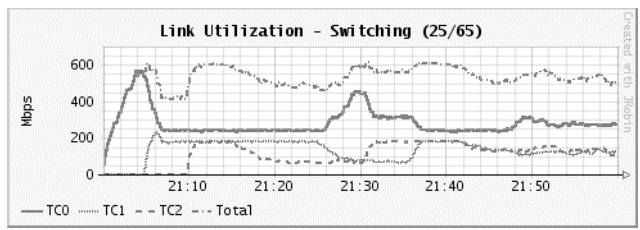

Fig. 9. Link 0 "utilization" with dynamic switching configured as 25/65

As a final remark, the utilization of operational parameters like the adopted for the switching of BAM behaviors (preemption/ utilization) through G-BAM configuration enables a managed BAMs evolution in such a way that it can be adjusted dynamically to fulfill the objective defined by the network administrator.

## VII. FINAL CONSIDERATIONS

The applicability of autonomy with Bandwidth Allocation Models (BAMs) was evaluated by considering BAM switching among distinct "behaviors" (RDM - MAM) using a HARD switching approach. The switching is executed in the simulation by configuring "BAM behaviors" (MAM and/or RDM) with the G-BAM model. The simulation over the NTT network demonstrate that, firstly, there are various possible configuration approaches that can be considered by a framework with autonomic characteristics and, secondly, that improvement in network quality parameters like link utilization and preemption are achievable. In general, the results point to the direction where some preemption and link utilization control is effectively achievable by using an autonomic framework approach and considering high-level management parameters.

In general, authors argue that autonomy and BAM-based dynamic resource allocation can coexist and may result in an overall improvement in network operation.

ACKNOWLEDGEMENTS

We gratefully acknowledge the financial support from FAPESB (Fundo de Amparo à Pesquisa do Estado da Bahia) – Brazil.